\documentclass[12pt,twoside]{article}

\usepackage{amsmath}
\usepackage{amsthm}
\usepackage{amssymb}
\usepackage{amscd}
\usepackage[british]{babel}
\usepackage{graphicx}
\usepackage{psfrag}
\usepackage{epsfig}
\usepackage{rotating}
\usepackage{times}

\title{Gravitational particle production in bouncing cosmologies}

\author{Jaume Haro $^a$ and Emilio Elizalde $^b$}

\date{April 2015}

\theoremstyle{plain}

\newcommand{\boxend}{\flushright{$\Box$}}

\newcommand{\R}{{\mathbb R}}               

\begin{document}

\maketitle

\noindent
\hspace*{-4mm} $^a$ Departament de Matem\`atica Aplicada I,
Universitat Polit\`ecnica de Catalunya
\newline
Diagonal 647, 08028 Barcelona, Spain
\newline
e-mail:  jaime.haro@upc.edu

\noindent
\hspace*{-4mm} $^b$ Consejo Superior de Investigaciones Científicas, ICE-CSIC and IEEC
\newline
UAB Campus, 08193 Bellaterra, Barcelona (Spain)
\newline
e-mail:  elizalde@ieec.uab.es

\thispagestyle{empty}

\begin{abstract}
It is argued that the Universe reheating in bouncing cosmologies could be explained via gravitational particle production,
as due to a sudden phase transition in the contracting regime. To this end, it is shown that, in the context of Loop Quantum Cosmology, gravitational production of massive particles
conformally
coupled with gravity in a matter-ekpyrotic bouncing Universe, where the sudden phase transition occurs in the contracting regime, yields a reheating
temperature which is in good agreement with cosmological observations.

\end{abstract}
\vspace{1cm}


\vspace{0.5cm}





\markboth{Gravitational particle production in bouncing  cosmologies}
{J. Haro and E. Elizalde}

\section{Introduction}
 The issue of  the Universe reheating in the matter bounce scenario (for descriptions of this scenario, see \cite{MBS}), via gravitational particle production
of light particles minimally coupled  and nearly conformally coupled with gravity, has been
recently addressed, respectively  in \cite{Quintin} and \cite{HARO}.
The idea is quite simple: to get efficient reheating one needs a non-adiabatic  transition between two different phases, in order to obtain enough  gravitational particle creation.
This process is usually called {\it preheating}.
At that time the Universe is far from thermal equilibrium, but then the created particles decay into very light particles which interact among themselves, thus producing a relativistic
plasma in thermal equilibrium and evolving like radiation. Finally, when the energy density of this relativistic plasma starts to dominate that of the background,
the Universe is reheated and enters  in a radiation dominated phase which matches well with the corresponding one for the hot Friedmann Universe.

In the matter bounce scenario the non-adiabatic  transition could be produced in the contracting phase. In fact, a transition from matter-domination
(this phase is essential in order to guarantee scale invariance of the power spectrum of perturbations, because modes that leave the Hubble radius during this phase have a
flat power spectrum \cite{bounce}),
to an ekpyrotic phase with equation of state $P=\omega\rho$ where $\omega>1$  could be assumed in the contracting regime. The  model thus obtained is being called in the
literature a matter-ekpyrotic bounce scenario  \cite{caiewing}, and since in the ekpyrotic phase the energy density of the field, namely $\varphi$,
evolves like $\rho_{\varphi}\sim a^{-3(1+\omega)}$, which in  the contracting phase increases faster than  $a^{-6}$, anisotropies do become negligible.  (Note that the energy density
of the anisotropies grows in the contracting phase as $a^{-6}$, that is,  faster than the matter energy density, and thus, without an ekpyrotic transition the isotropy of the bounce
would be destroyed; this is the so-called Belinsky-Khalatnikov-Lifshitz instability \cite{lifshitz}.) Moreover, the field energy density  also grows faster than that of the particles
produced, what means that the field dominates the Universe evolution  in the contracting phase. But, when the Universe bounces, the energy density of the
relativistic plasma generated by the decay of the particles created due to the phase transition will eventually dominate, and thus,
the Universe will finally become radiation-dominated, in a way which may perfectly match the results for the hot Friedmann Universe.

This situation is similar to the one occurring in inflationary cosmology. To wit,  for inflationary  models with potentials not having a minimum (the so-called non-oscillatory models \cite{fkl}),
there is an abrupt transition from a quasi de Sitter regime to a kinetic-dominated period.
During this transition, massless, nearly conformally coupled particles are created which energy density evolves like $\rho_r\sim a^{-4}$. On the other hand, when the inflaton field,
$\Phi$, enters a kinetic-dominated period its  energy density behaves like $\rho_{\Phi}\sim a^{-6}$ \cite{ford, peebles, spokoiny}, what means that the inflation energy density
decreases faster than that of the radiated particles, and thus, the Universe will become radiation dominated, to eventually match the hot Friedmann Universe.
On the other hand, in inflationary models with potentials which do exhibit a minimum,
adiabaticity is broken when the inflaton field oscillates \cite{linde}; at that moment massive particles are created and, after the preheating stage, they decay into light
particles which thermalize and whose energy density eventually dominates the one of the inflaton. Therefore, the Universe becomes radiation dominated, too.

In the present analysis we will consider, in flat Friedmann-Lema{\^\i}tre-Robertson-Walker (FLRW) spacetime, the matter-ekpyrotic bounce scenario in the context of
Loop Quantum Cosmology \cite{wilson}, where a phase transition to the ekpyrotic phase is produced in the
contracting regime. One of the most remarkable properties of this model is its simplicity, in the sense that in the non-singular bounce the adiabacity is not broken, since
it is well-known that given the Equation of State $P=w\rho$,
dealing with Loop Quantum Cosmology, the scale factor evolves  as \cite{wilson1}
$a(t)=\left(\frac{3\rho_c}{4M_{Pl}^2}(1+w)^2t^2+1\right)^{\frac{1}{3(1+w)}}$, where $\rho_c$ is  the energy density at bouncing time and $M_{Pl}$ the Planck mass. 
One can easily see that, at bouncing time ($t=0$),
the scale factor is an smooth function what preserves the adiabacity.
This situation does not happen---i.e., the transition from contraction to expansion is not adiabatic---in other bouncing cosmologies; for example,  in cosmologies within
the framework of General Relativity where the violation of the null energy
condition at bouncing time (needed in order to have a bounce in the flat FLRW geometry), could be possible by incorporating new forms of matter, such as phantom \cite{aw}
or quintom fields \cite{cqplz07}, Galileons \cite{qeclz}, or phantom condensates \cite{lbp}.

In our simple study, we also deal with heavy
massive particles conformally coupled to gravity, because in Fourier space, these particles can be depicted as a set of harmonic oscillators whose
positive frequency is time dependent. This allows to easily interpret the number of particles created and its energy density as the $00$ component
of the stress-energy tensor---where the vacuum zero point energy of the set of oscillators must be subtracted, in order to get a well-defined quantity (see, e.g., \cite{ee_b2}). Then,
the energy density of the
created particles will be calculated and, in the expanding phase, when it is of the same order as the background energy, the Universe will get reheated, with a temperature of the
order of this energy density power $1/4$. We will see that the reheating temperature will basically depend on the energy density of the background when the sudden phase
transition is produced, and thus, by properly choosing this energy density, the reheating temperature will be shown to match well that of astronomical observations.


Units used in the work are: $\hbar=c=1$.

\section{Particle production of massive conformally coupled particles}
 Let $\phi\equiv\frac{\chi}{a}$ be a  massive scalar field conformally coupled with gravity, $a$ being the scale factor. In Fourier space, the
 corresponding Klein-Gordon equation, in the FLRW spacetime,
 is given by the following set of harmonic oscillators, with time dependent frequency \cite{haro1}:
\begin{eqnarray}\label{KG}\chi_k''+\omega_k^2(\eta)\chi_k=0,
\end{eqnarray}
where $\omega_k^2(\eta)=|k|^2+m^2a^2(\eta)$, being $\eta$ the conformal time and $m$ the mass of the field.

Using the instantaneous diagonalization method \cite{Instantaneous}, the square of the $\beta$-Bogoliubov coefficient is given by
\begin{eqnarray}\label{bogoliubov}
 |\beta_k(\eta)|^2=\frac{1}{\omega_k(\eta)}\left[\frac{1}{2}
 \left(|\chi_k'(\eta)|^2+\omega^2_k(\eta)|\chi_k(\eta)|^2
 \right) -\frac{1}{2}\omega_k(\eta)\right].
\end{eqnarray}
We assume that at some given time, $\eta_E$, there is a sudden phase transition between two adiabatic phases. Note that, as $\frac{d^s\omega_k}{d\eta^s}\sim H^sm$ ($H$ being the Hubble 
parameter),
if one
assumes $H\leq H_{max}\ll m$ (i.e., the generated particles have a large mass), where $H_{max}$ is the maximum value of
the Hubble parameter---what always happens in holonomy corrected Loop Quantum Cosmology because in the plane $(H,\rho)$ the modified Friedmann equation depicts an ellipse 
that is a bounded 
curve \cite{aho}---then, for
$n=-1+(r+1)\alpha+ (s+1)\beta$ with $\alpha+\beta-1>0$, one gets
\begin{eqnarray}
 \frac{d^n\omega_k}{d\eta^n}\ll \left( \frac{d^r\omega_k}{d\eta^r}\right)^{\alpha}\left(\frac{d^s\omega_k}{d\eta^s} \right)^{\beta},
\end{eqnarray}
in particular, for $n=1$, $\alpha=2$ and $r=s=\beta=0$, one has $\omega_k'\ll \omega_k^2$.

During the adiabatic regimes we use the first order WBK solution \cite{w05} of (\ref{KG})
\begin{eqnarray}\label{wkb}
 \chi_{1,k}(\eta)=\frac{1}{\sqrt{2W_{1,k}(\eta)}}e^{-i\int^{\eta}W_{1,k}(\eta)d\eta}\mbox{ where }
 W_{1,k}=\omega_k-\frac{1}{2\omega_k}\left[\frac{\omega_k''}{2\omega_k}-\frac{3}{4}\left(\frac{\omega_k'}{\omega_k}\right)^2  \right],
\end{eqnarray}
together with its conjugate, to approximate the mode solutions.

In the first phase, the mode corresponding to the vacuum state is approximated by $\chi_{1,k}$. Before the sudden phase transition this mode becomes
$a_k\chi_{1,k}+b_k\chi_{1,k}^*$, with
\begin{eqnarray}
 b_k=-i{\mathcal W}[\chi_{1,k}(\eta_E^-); \chi_{1,k}(\eta_E^+)],
\end{eqnarray}
where ${\mathcal W} $ is the Wronskian and  $\chi_{1,k}(\eta_E^{-})$ (resp.  $\chi_{1,k}(\eta_E^{+})$) is the value of the mode just before (after) the time $\eta_E$.
A simple calculation using  the relation $|a_k|^2-|b_k|^2=1$, yields
\begin{eqnarray}
 |\beta_k(\eta)|^2=\frac{1}{2\omega_k(\eta)}\left[|\chi_{1,k}'(\eta)|^2+\omega^2_k(\eta)|\chi_{1,k}(\eta)|^2
 \right] -\frac{1}{2}+\frac{|b_k|^2}{\omega_k(\eta)}\left[|\chi_{1,k}'(\eta)|^2+\omega^2_k(\eta)|\chi_{1,k}(\eta)|^2
 \right]\nonumber\\
 +\frac{1}{\omega_k(\eta)}{\mathcal Re}\left(a_kb^*_k[(\chi_{1,k}'(\eta))^2+\omega^2_k(\eta)(\chi_{1,k}(\eta))^2]  \right)=|b_k|^2+{\mathcal O}
 \left(\frac{\omega'_k(\eta)}{\omega^2_k(\eta)} \right)\cong |b_k|^2.
\end{eqnarray}
Assuming that the derivative of $a$ has a discontinuity at some time value, $\eta=\eta_E$ (namely that the
Hubble parameter is discontinuous at $\eta=\eta_E$), one gets
\begin{eqnarray}\label{A}
 |\beta_k(\eta)|^2\cong \frac{(\omega_k'(\eta_E^+)-\omega_k'(\eta_E^-))^2}{16\omega^4_k(\eta_E)},
\end{eqnarray}
which, in terms of the cosmic time,  reads
\begin{eqnarray}\label{B}
 |\beta_k(t)|^2\cong \frac{m^4 a_E^6(H_E^+-H_E^-)^2}{16(|k|^2+ m^2 a_E^2)^3},
\end{eqnarray}
where $a_E\equiv a(t_E)$ and  $H_E^{\pm}\equiv H(t_E^{\pm})$.

 When the Hubble parameter is continuous but its derivative is discontinuous, that is, the scalar curvature is discontinuous, at some instant $t_E$, a similar calculation leads to
 \begin{eqnarray}\label{C}
 |\beta_k(t)|^2\cong \frac{m^4 a_E^8(R_E^+-R_E^-)^2}{2304(|k|^2+ m^2 a_E^2)^4}=\frac{m^4 a_E^8(\dot{H}_E^+-\dot{H}_E^-)^2}{64(|k|^2+ m^2 a_E^2)^4},
\end{eqnarray}
where $R(t)$ is the scalar curvature and $R_E^{\pm}\equiv R(t_E^{\pm})$.

On the other hand, the density of produced particles and their corresponding energy density are respectively given by \cite{bd82}
\begin{eqnarray}
 N_{\chi}(\eta)=\frac{1}{(2\pi a(\eta))^3}\int_{\R^3}|\beta_k(\eta)|^2 d^3 k; \quad \rho_{\chi}(\eta)=\frac{1}{(2\pi a(\eta))^3a(\eta)}\int_{\R^3}\omega_k(\eta)|\beta_k(\eta)|^2 d^3 k.
\end{eqnarray}
Note that, from the definition of the $\beta$-Bogoliubov coefficients (\ref{bogoliubov}), we can see that the energy density $\rho(t)$ is the $00$ component of the stress-energy 
tensor where, in order to obtain a convergent quantity, one needs to subtract the vacuum zero-point energy \cite{ee_b2}.

As a consequence, in the case when the Hubble parameter has a discontinuity at $t=t_E$, from formula (\ref{B}) one obtains, in cosmic time,
\begin{eqnarray}\label{formula1}
 N_{\chi}(t)=\frac{m(H^+_E-H_E^-)^2}{512 \pi}\left(\frac{a_E}{a(t)}\right)^3;
\quad \rho_{\chi}(t)=
\frac{m^2(H^+_E-H_E^-)^2}{32 \pi^2}\left(\frac{a_E}{a(t)}\right)^4
I_3\left(\frac{a(t)}{a_E}\right),
\end{eqnarray}
where
\begin{eqnarray}
 I_3\left(\frac{a(t)}{a_E}\right)\equiv\int_0^{\infty}\frac{x^2\sqrt{x^2+\left(\frac{a(t)}{a_E}\right)^2}}{(x^2+1)^3}dx=
 \left\{\begin{array}{ccc}
   \frac{\left(\frac{a(t)}{a_E}\right)^4\arctan\left(\sqrt{\left(\frac{a(t)}{a_E}\right)^2-1}\right)}{8\left[\left(\frac{a(t)}{a_E}\right)^2-1\right]
   \sqrt{\left(\frac{a(t)}{a_E}\right)^2-1}} +\frac{\left(\frac{a(t)}{a_E}\right)^2-2}{8\left[\left(\frac{a(t)}{a_E}\right)^2-1\right]},& \mbox{for}&      \frac{a(t)}{a_E} >1,\\
   1/3,& \mbox{for}&       \frac{a(t)}{a_E}=1,\\
   \frac{\left(\frac{a(t)}{a_E}\right)^4\tanh^{-1}\left(\sqrt{1-\left(\frac{a(t)}{a_E}\right)^2}\right)}{8\left[\left(\frac{a(t)}{a_E}\right)^2-1\right]
   \sqrt{1-\left(\frac{a(t)}{a_E}\right)^2}} +\frac{\left(\frac{a(t)}{a_E}\right)^2-2}{8\left[\left(\frac{a(t)}{a_E}\right)^2-1\right]},& \mbox{for}&      \frac{a(t)}{a_E} <1.
                      \end{array}\right.
\end{eqnarray}
From this result it follows that, when $\frac{a(t)}{a_E}\ll 1$,
$$\rho_{\chi}(t)\cong
\frac{m^2(H^+_E-H_E^-)^2}{128\pi^2}\left(\frac{a(t)}{a_E}\right)^4
$$
and, consequently, if the sudden phase transition takes place in the contracting phase, then
the generated particles will evolve, in a part of the contracting and expanding phase, exactly like radiation does. However, if these particles do not decay as a relativistic plasma then,
when $\frac{a(t)}{a_E}\gg 1$, they will evolve as matter since, in this case
$$\rho_{\chi}(t)\cong
m N_{\chi}(t)= \frac{m(H^+_E-H_E^-)^2}{512 \pi}\left(\frac{a_E}{a(t)}\right)^3.
$$

The same happens  when the Hubble parameter is continuous but the scalar curvature  has a discontinuity at $t=t_E$. In such case, from formula (\ref{C}), one obtains, in cosmic time,
\begin{eqnarray}
 N_{\chi}(t)=\frac{(\dot{H}^+_E-\dot{H}_E^-)^2}{4096 m\pi}\left(\frac{a_E}{a(t)}\right)^3; \quad \rho_{\chi}(t)=
 \frac{(\dot{H}^+_E-\dot{H}_E^-)^2}{128 \pi^2}\left(\frac{a_E}{a(t)}\right)^4
 I_4\left(\frac{a(t)}{a_E}\right),
\end{eqnarray}
where $I_4\left(\frac{a(t)}{a_E}\right)\equiv \int_0^{\infty}\frac{x^2\sqrt{x^2+\left(\frac{a(t)}{a_E}\right)^2}}{(x^2+1)^4}dx $, is related with
$I_3\left(\frac{a(t)}{a_E}\right) $ through the formula
$$I_4(y)=\frac{1}{6y}\frac{d(y^2I_3(y))}{dy}.
$$
In the asymptotic case $\frac{a(t)}{a_E}\ll 1$, one has
$$\rho_{\chi}(t)\cong \frac{(\dot{H}^+_E-\dot{H}_E^-)^2}{1536\pi^2}\left(\frac{a_E}{a(t)}\right)^4$$
and, when $\frac{a(t)}{a_E}\gg 1$,
$$\rho_{\chi}(t)\cong mN_{\chi}(t)=\frac{(\dot{H}^+_E-\dot{H}_E^-)^2}{4096\pi}\left(\frac{a_E}{a(t)}\right)^3.$$
Finally, it is important to stress that,  when the Hubble parameter is continuous but the curvature is discontinuous, the energy density does not depend on the mass of the field.

\section{Bouncing models}
In this Section
we will study two simple bouncing models coming from Loop Quantum Cosmology, where holonomy
corrections modify Friedmann's equation as follows \cite{lqc}
\begin{eqnarray}\label{FriedmannLQC}
H^2=\frac{\rho}{3M_{Pl}^2}\left(1-\frac{\rho}{\rho_c} \right),
\end{eqnarray}
 being $M_{Pl}$  the reduced Planck mass and $\rho_c$  the so-called {\it critical energy density}, the value of the energy density at the bouncing time, its maximal value.
 From this equation it is clear
 that, when $\rho\ll \rho_c$, one recovers the standard Friedmann equation, that is, holonomy corrections can then be disregarded. Moreover, we can see that Eq.~(\ref{FriedmannLQC}) corresponds to an ellipse in the plane $(H,\rho)$ \cite{aho}.

 As we have explained in the Introduction, holonomy corrected Loop Quantum Cosmology provides a non-singular bounce that preserves adiabacity. In fact, if during the
 transition from contraction to expansion the Universe is dominated by a barotropic fluid with Equation of State $P=w\rho$, the solution of  (\ref{FriedmannLQC}) will be given by
 \cite{wilson1}
 \begin{eqnarray}
  a(t)=\left(\frac{3\rho_c}{4M_{Pl}^2}(1+w)^2t^2+1\right)^{\frac{1}{3(1+w)}},
 \end{eqnarray}
which has an smooth behavior at the bouncing time $t=0$, and thus, leads to an adiabatic transition.

In both models we will assume that, in the contracting phase, there is a sudden transition from the matter domination phase to the ekpyrotic one, that breaks adiabacity. And since,
as we have already seen, holonomy corrected Loop Quantum Cosmology preserves adiabacity at the bouncing time, in both models all particles are created, in the contracting regime, during this sudden
transition.

\subsection{First model}
In this model we will consider
 a discontinuous Hubble parameter  at $t=t_E$ which scale factor is given by
\begin{eqnarray}
 a(t)=\left\{\begin{array}{ccc}
  a_E\left(\frac{t}{t_E}\right)^{2/3},& \mbox{when} & t<t_E,\\
  \left(\frac{3\rho_c}{4M_{Pl}^2}(1+w)^2t^2+1\right)^{\frac{1}{3(1+w)}},& \mbox{when} & t\geq t_E,
             \end{array}\right.
\end{eqnarray}
where we assume $w>1$ in order to have an ekpyrotic phase.

Note that for $t<t_E$, the scale factor is the solution of the standard Friedmann equation $H^2=\frac{\rho}{3M_{Pl}^2}$, for a matter dominated Universe ($P=0$), and for
$t>t_E$, the scale factor is the solution of the holonomy corrected Friedmann equation (\ref{FriedmannLQC}), for an ekpyrotic Universe whose  EoS is $P=w\rho$.
Assuming moreover that the sudden transition is produced at very early times, $\frac{3\rho_c}{4M_{Pl}^2}(1+w)^2t^2 \gg 1$, one obtains
$H^+_E-H_E^-=-\frac{w}{1+w}H^-_E$, and thus, from Eq.~(\ref{formula1}),  the density of generated particles reads
\begin{eqnarray}
 N_{\chi}(t)=\frac{mw^2(H^-_E)^2}{512 \pi(1+w)^2}\left(\frac{a_E}{a(t)}\right)^3.
\end{eqnarray}

Now we will show that this amount of produced particles is enough to reheat our Universe.
To prove this statement note, first of all, that immediately after the transition, the energy density of the produced particles and the background energy density are, respectively,
\begin{eqnarray}
 \rho_{\chi}(t_E^+)\cong
 \frac{m^2w^2\rho_E^-}{96 \pi M_{Pl}^2(1+w)^2}, \quad \mbox{and}\quad \rho(t_E^+)=\frac{\rho_E^-}{(1+w)^2},
\end{eqnarray}
where $ \rho_E^-$ is the background energy density at the end of the matter domination phase.

Then,
in order to have a sub-dominant energy density for the  particles created at the beginning of the ekpyrotic phase---so that back-reaction effects can be disregarded---one has to assume 
that $M_{Pl}\gg mw$. Since in the contracting phase the background energy density increases
as $\left(\frac{a_E}{a(t)}\right)^{3(1+w)}$, it will be dominant during the whole phase, and
only in the expanding one will both energy densities be of the same order, at some time $t=t_R$. But  then, the order of the reheating temperature may be obtained as
$T_R\sim \rho_{\chi}^{1/4}(t_R)$.

Secondly,
as we have already explained in the Introduction, the particles created are far from being in thermal equilibrium. The more massives particles will decay into lighter particles, which will interact
through multiple scattering, thus redistributing their energies until they yield a relativistic plasma in thermal equilibrium (see \cite{abc} for a very detailed description of this process).

Let $\Gamma$ be the decay rate of massive  $\chi$-particles. The decay will be accomplished at time $t_{dec}$ when $\Gamma |t_{dec}-t_E|\cong 1$. If, in order to simplify things
(even though it is not essential), we assume this happens in the contracting phase
when holonomy corrections can be neglected, then it turns out that, as
for $t>t_E$ one has $H\cong \frac{2}{3(1+w)t}$, one gets
\begin{eqnarray}
 \frac{2}{3(1+w)}\frac{H_E^+-H_{dec}}{H_E^+H_{dec}}\cong \Gamma^{-1}\Longrightarrow H_{dec}\cong \frac{2\Gamma H_E^+}{2\Gamma+{3(1+w)}H^+_E }.\end{eqnarray}
 This means that the decay rate must satisfy the constrain
 \begin{eqnarray}
  0<|H_{dec}|\ll \sqrt{\frac{\rho_c}{M_{Pl}^2}}\Longrightarrow \Gamma\gg -\frac{3}{2}(1+w)\frac{H_E^+}{\sqrt{\rho_c}+M_{Pl}H_E^+},
 \end{eqnarray}
because we are assuming that the decay is produced, in the contracting phase, whenever holonomy corrections can be disregarded.

 When the decay is effective, the background energy density reads
 \begin{eqnarray}\rho_{dec}\cong 3M_{pl}^2
 \left(\frac{2\Gamma H_E^+}{2\Gamma+{3(1+w)}H^+_E }\right)^2\ll \rho_c.
\end{eqnarray}
And thus, once the decay has taken place, a re-distribution
of energies among the different particles occurs---{\it kinetic equilibrium}---and, also,
an increase in the number of particles---{\it chemical equilibrium}. That is to say, in such process, in order to obtain a relativistic plasma in thermal equilibrium,  
both number-conservating and number-violating reactions are definitely involved.

Finally, to calculate the reheating temperature, one has to impose that both the energy densities of the relativistic plasma and of the background are of the same 
order. This will happen, in the expanding phase, when $a(t_R)\gg a_E$, namely
\begin{eqnarray}
 \rho(t_R)= \frac{\rho_E^-}{(1+w)^2}\left(\frac{a_E}{a(t_R)}\right)^{3(1+w)}\sim \rho_{\chi}(t_R)\cong \frac{m^2w^2\rho_E^-}{M_{Pl}^2(1+w)^2}\left(\frac{a_E}{a(t_R)}\right)^{4},
\end{eqnarray}
what means that
\begin{eqnarray}
 \frac{a_E}{a(t_R)}\sim \left(\frac{m^2w^2}{M_{Pl}^2} \right)^{\frac{1}{3w-1}}.
\end{eqnarray}
And thus,  the reheating temperature will become
\begin{eqnarray}
 T_R\sim \rho_{\chi}^{1/4}(t_R)\sim \left[\frac{\rho_E^-}{M_{Pl}^4(1+w)^2}\left(\frac{m^2w^2}{M_{Pl}^2} \right)^{\frac{3(w+1)}{3w-1}}\right]^{1/4} M_{Pl}.
\end{eqnarray}

Note that in the asymptotic case $w\gg 1$, the reheating temperature has this simple expression
\begin{eqnarray}
 T_R\sim \left(\frac{m^2\rho_E^-}{M_{Pl}^6}\right)^{1/4} M_{Pl},
\end{eqnarray}
and in this case, assuming from recent observations that $T_R\sim 10^{-7}M_{Pl}$  (see \cite{Quintin}), one concludes that the energy density of the background at the end of the
matter domination stage has to be necessarily of the order
$$\rho_E^-\sim \left(10^{-28}\frac{M^2_{Pl}}{m^2} \right) M_{Pl}^4.$$
That is, our model will  definitely match observations, namely, the amount of created particles is enough to reheat the Universe,  when the phase transition is produced at very low energy densities, or equivalently, at very early times in the contracting phase.

To finish this example, it is instructive to compare it with its inflationary dual, that is, a Universe characterized by a sudden transition from the de Sitter phase to a radiation-dominated one.
Suppose inflation is produced by a false vacuum energy density  $\rho_E$, then the scale factor is given by \cite{ford}
\begin{eqnarray}
 a(\eta)=\left\{\begin{array}{ccc}
                \frac{1}{\eta H_E}& \mbox{for}& \eta<\eta_E<0\\
    H_E(\eta-\eta_E)+ \frac{1}{\eta_E H_E}  & \mbox{for}& \eta>\eta_E,
               \end{array}\right.
\end{eqnarray}
where $H_E$ is the value of the Hubble parameter during the de Sitter phase. Then,
for  heavy massive ($m\gg H_E$) particles conformally coupled with gravity, since $H(\eta)$ is discontinuous at $\eta=\eta_E$, one can use Eq.~(\ref{formula1}) to obtain
$\rho_{\chi}(t_E)\sim {H_E^2}{m^2}$. Thus, after the decay into light particles and thermalization, the reheating temperature, as in the matter-ekpyrotic bounce scenario
with $w\gg 1$, is given by
\begin{eqnarray}
 T_R\sim ({H_E}{m})^{1/2}=\left(\frac{m^2\rho_E}{M_{Pl}^6}\right)^{1/4} M_{Pl},
\end{eqnarray}
which is another manifestation of the existing duality between the de Sitter regime in the expanding phase and matter-domination in the contracting one, pointed out, for the first time,
in \cite{wands}.

\subsection{Second model}

As second model of the matter-ekpyrotic bouncing scenario in Loop Quantum Cosmology, we will consider a sudden transition where the Hubble parameter is continuous but the scalar curvature 
has a discontinuity, at $t=t_E$. 
If one considers Eq.~(\ref{FriedmannLQC}) with and EoS of the form
\begin{eqnarray}
 P(\rho)=\left\{\begin{array}{ccc}
              0&\mbox{when}& \rho<\rho_E\\
              w\rho& \mbox{when}& \rho>\rho_E,
                \end{array}
\right.
\end{eqnarray}
in the contracting phase, and $P(\rho)=w\rho$ in the expanding one, the
 scale factor will be given by
\begin{eqnarray}
 a(t)=\left\{\begin{array}{ccc}
  a_E\left(\frac{t-t_0}{t_E-t_0}\right)^{2/3},& \mbox{when} & t<t_E,\\
  \left(\frac{3\rho_c}{4M_{Pl}^2}(1+w)^2t^2+1\right)^{\frac{1}{3(1+w)}},& \mbox{when} & t\geq t_E,
             \end{array}\right.
\end{eqnarray}
where $t_0\equiv t_E-\frac{2}{3H_E}$, being $H_E=\sqrt{\frac{\rho_E}{3M_{Pl}^2}}$ the value of the Hubble parameter at the time of the phase transition.
Note that, since we are assuming that the transition is produced at very early times, during matter domination, holonomy corrections could safely be disregarded, and thus, 
the exact value of the scale factor $a(t)$ is approximately given by $a_E\left(\frac{t-t_0}{t_E-t_0}\right)^{2/3}$.

In this case, $R^+_E-R_E^-=-9wH_E^2$, and then the density of produced particles is
\begin{eqnarray}
 N_{\chi}(t)=\frac{81w^2H_E^4}{4096 m \pi}\left(\frac{a_E}{a(t)}\right)^3.
\end{eqnarray}
On the other hand, at the transition time the energy density of these generated particles evolves as
\begin{eqnarray}
 \rho_{\chi}(t)\sim
 \frac{w^2\rho_E^2}{M_{Pl}^4}.
\end{eqnarray}
As a consequence, such energy density will be sub-dominant at the beginning of the ekpyrotic phase, provided $w^2\rho_E\ll {M_{Pl}^4}$. And since the background energy density increases fastly in the
contracting regime and also decreases fastly in the expanding one, the Universe will eventually reheat, in the expanding phase, when both these energy densities are of the same order. This will happen for $a(t_R)\gg a_E$.

Imposing this condition, one obtains
\begin{eqnarray}
 \frac{a_E}{a(t_R)}\sim \left(\frac{w^2\rho_E}{M_{Pl}^4} \right)^{\frac{1}{3w-1}},
\end{eqnarray}
and thus, the reheating temperature will be
\begin{eqnarray}
 T_R\sim \rho_{\chi}^{1/4}(t_R)\sim \left[\frac{\rho_E}{M_{Pl}^4}\left(\frac{\rho_Ew^2}{M_{Pl}^4} \right)^{\frac{3(w+1)}{3w-1}}\right]^{1/4} M_{Pl}.
\end{eqnarray}
Finally, note that in the
asymptotic case, $w\gg 1$, the reheating temperature acquires the simple form
\begin{eqnarray}
 T_R\sim \left(\frac{w\rho_E}{M_{Pl}^4}\right)^{1/2} M_{Pl}
\end{eqnarray}
and, in this case, assuming $T_R\sim 10^{-7}M_{Pl}$, it turns out that, in order to match with recent observational data, the energy density of the background  at the phase transition must be of the order
$$\rho_E\sim 10^{-14}\frac{1}{w} M_{Pl}^4.$$

\section{Some remarks on massless particle production}
When one deals with  particles non-conformally coupled with gravity, the diagonalization method results, in general, in a divergent number of produced particles. This means that, in
such case, the concept of particles created has to be defined in a different way. We believe that, in the case of massless particles non-conformally coupled with gravity, the
modes associated to the vacuum must be those which, at an early time, satisfy the asymptotic condition
$\chi_k(\eta)\rightarrow \frac{e^{-i|k|\eta}}{\sqrt{2|k|}}$.

Since the  Klein-Gordon equation for massless particles reads
\begin{eqnarray}
 \chi''_k+\left(|k|^2-(1-6\xi)\frac{a''}{a}  \right)\chi_k=0,
\end{eqnarray}
where $\xi$ is the coupling constant (equal to $1/6$ when the coupling is conformal),
with our definition,  the modes that
define the vacuum state in the matter dominated  phase are
\begin{eqnarray}\label{a47}
 \chi_{ k}^{matt}(\eta)=e^{-i(\frac{\pi\nu_{matt }}{2}+\frac{\pi}{4})}\sqrt{\frac{\pi\eta}{4}}H^{(2)}_{\nu_{matt}}({ |k|} \eta),
\end{eqnarray}
where $H^{(2)}_{\nu_{matt}}$ is Hankel's function \cite{as72},
with $\nu_{matt}\equiv\sqrt{\frac{9}{4}-12\xi}$. Here, we will assume that $\xi<\frac{3}{16}$, in order that $\nu_{matt}$ becomes a real number.
Note that these modes are the same which define the vacuum in the de Sitter regime in the expanding phase because, in both cases, $\frac{a''}{a}=\frac{2}{\eta}$; this is 
the duality beetwen matter-domination during contraction and de Sitter phase during expansion \cite{wands}.

On the other hand, if the phase transition takes place at very early times, when holonomy corrections can be safely disregarded, then the modes which define the vacuum state in the ekpyrotic
regime, just following the phase transition, are given by
\begin{eqnarray}\label{a48}
 \chi_{ k}^{ekpy}(\eta)=e^{-i(\frac{\pi\nu_{ekpy}}{2}+\frac{\pi}{4})}
 \sqrt{\frac{\pi\eta}{4}}H^{(2)}_{\nu_{ekpy}}({ |k|} \eta),
\end{eqnarray}
with $\nu_{ekpy}\equiv\sqrt{\frac{1}{4}+\frac{2(1-3w)}{(1+3w)^2}(1-6\xi)}$. Consequently, the squared modulus of the beta Bogoliubov coefficients are calculated as
\begin{eqnarray}\label{wronskia}
 |\beta_k|^2=\left|{\mathcal W}\left [\sqrt{\frac{\pi\eta_E}{4}}H^{(2)}_{\nu_{matt}}({ |k|} \eta_E);
 \sqrt{\frac{\pi\eta_E}{4}}H^{(2)}_{\nu_{ekpy}}({ |k|} \eta_E)\right]\right|^2.
\end{eqnarray}

What is remarkable is that for $w\gg 1$ one gets $\nu_{ekpy}\cong \frac{1}{2}$, and thus,
$\chi_{ k}^{ekpy}(\eta)\cong\frac{e^{-i|k|\eta}}{\sqrt{2|k|}}$. That is, when $w\gg 1$ the phase transition from the matter to the ekpyrotic phase in the contracting regime is exactly
the same as the phase transition from the de Sitter to a radiation dominated phase in the expanding one. This last phase transition has been studied in several works, in order
to explain the reheating in inflationary models via gravitational particle production (see, e.g., \cite{ford, dv, gio, haro1}).

Then, for $w\gg 1$, when one consider modes well outside of the Hubble radius, i.e., which satisfy $|k\eta_E|\ll 1$, we can use the formula for small arguments \cite{as72}
\begin{eqnarray}\label{b49}
 H^{(2)}_{\nu}(z)\cong \frac{i}{\pi}\left(z/2\right)^{-\nu}\Gamma(\nu)-\frac{ie^{i\pi\nu}}{\sin(\pi\nu)}\left(z/2\right)^{\nu}
\frac{1}{\Gamma(\nu+1)},
\end{eqnarray}
to obtain, after matching at $\eta_E$ with the mode $\alpha_k \frac{e^{-i|k|\eta}}{\sqrt{2|k|}}+\beta_k \frac{e^{i|k|\eta}}{\sqrt{2|k|}}$, or using Eq.~(\ref{wronskia}), the formula
\begin{eqnarray}\label{a49}
 |\beta_{ k}|^2\cong \frac{1}{16\pi}\left(|
 k\eta_E|/2\right)^{-2\nu_{matt}-1}\Gamma^2(\nu_{matt})\left(1/2-\nu_{matt}\right)^2.
\end{eqnarray}
And in the opposite case, $|k\eta_E|\gg 1$---i.e., for models well inside the Hubble radius---from the asymptotic expression for large arguments \cite{as72}
\begin{eqnarray}\label{a50}
 H^{(2)}_{\nu}(z)\cong e^{i(\frac{\pi\nu}{2}+\frac{\pi}{4})}\sqrt{\frac{2}{\pi z}}
\left(1-i\frac{4\nu^2-1}{8z}\right)e^{-iz},
\end{eqnarray}
after matching at  $\eta_E$, one obtains
\begin{eqnarray}\label{a51}
 |\beta_{ k}|^2\cong \frac{(4\nu_{matt}^2-1)^2}{4{ |k|}^4\eta_E^4}.
\end{eqnarray}
It follows from these expressions that the energy density of the produced particles is always ultraviolet divergent, and also, that for $\nu_{matt}\geq 1$ (i.e., $\xi\leq \frac{5}{48}$) the
number of  particles created is infrared divergent, although their energy density is only infrared divergent for $\nu_{matt}\geq 3/2$, i.e., for $\xi\leq 0$, what includes
massless particles minimally coupled to gravity.

To avoid infrared divergencies one must assume that the coupling constant $\xi$ belongs in the range $(5/48,3/16)$ and, in order to have a
finite energy density one needs to assume, as was shown in \cite{ford}, that the scalar curvature is regular (continuous, at the very least) during the phase transition.

Simple formulas can be obtained if one only considers the production of particles whose modes leave the Hubble radius before the phase transition, i.e., for modes satisfying
$|k\eta_E|<1$. In this case, when $5/48<\xi<3/16\Longleftrightarrow 0<\nu_{matt}<1$ (to avoid infrared divergencies), using Eq.~(\ref{a49}) one gets
\begin{eqnarray}\label{a52}
 N_{\chi}(t)\cong \frac{4^{\nu_{matt}}}{256\pi^3(1-\nu_{matt})}\Gamma^2(\nu_{matt})\left(\frac{1}{2}-\nu_{matt}\right)^2{|H_E|^3}\left(\frac{a_E}{a(t)}\right)^3,
\end{eqnarray}
and
\begin{eqnarray}\label{a53}
 \rho_{\chi}(t)\cong \frac{4^{\nu_{matt}}}{256\pi^3(3-2\nu_{matt})}\Gamma^2(\nu_{matt})\left(\frac{1}{2}-\nu_{matt}\right)^2{H_E^4}\left(\frac{a_E}{a(t)}\right)^4.
\end{eqnarray}
Note that the case $\nu_{matt}=1/2\Longleftrightarrow \xi=\frac{1}{6}$ is the conformally
coupled one, and thus, there is not particle production there.

Finally, provided the phase transition is smooth enough, the energy density of the massless particles being produced will be of the order
$\rho_{\chi}(t)\sim \left(\frac{1}{2}-\nu_{matt}\right)^2\frac{\rho_E^2}{M_{Pl}^4}\left(\frac{a_E}{a(t)}\right)^4\sim \left(\frac{1}{2}-\nu_{matt}\right)^2\frac{\rho_E^2}{M_{Pl}^4}$, because
for $w\gg 1$ the approximation $\left(\frac{a_E}{a(t)}\right)^4\cong 1$ holds in the ekpyrotic case.
On its turn, the reheating temperature will be of the order
\begin{eqnarray}
 T_R\sim \left(\left|\frac{1}{2}-\nu_{matt}\right|\frac{\rho_E}{M_{Pl}^4} \right)^{1/2} M_{Pl},
\end{eqnarray}
which coincides with the reheating temperature obtained
in the transition from the de Sitter phase to a radiation-dominated one---because in this inflationary case
one has the well-known result \cite{peebles, haro1, dv} $\rho_{\chi}\sim \left(\frac{1}{2}-\nu\right)^2 H_E^4$ (compare with Eq. (\ref{a53})), with
$\nu=\nu_{matt}=\sqrt{\frac{9}{4}-12\xi}$.
This illustrates, once again, the duality existing between the transition form the de Sitter regime to a radiation-dominated Universe, in the expanding phase, and 
the transition from matter domination to an ekpyrotic regime, in the contracting one.

\section{Conclusions}
We have shown in this paper that, for  models arising in holonomy corrected Loop Quantum Cosmology, a simple phase transition which takes place in the contracting phase,
from matter domination to an ekpyrotic regime, can lead in fact to  the production 
of heavy particles conformally coupled with gravity
in a sufficient amount (see Eqs.~($17$) and  ($30$)) and with enough energy density, able
to reheat the Universe in the expanding phase.
This is quite a remarkable conclusion. Moreover, as a bonus we see that the expression of the reheating temperature is actually
quite simple (see formulae ($25$), ($34$) and ($45$)), what does not happens in other bouncing models where adiabaticity is broken at bouncing time
(see for instance, Eq.~$(60)$ of \cite{Quintin}, where  production of minimally coupled particles and reheating are studied in the two-fields matter bounce scenario). 
In our results, the reheating temperature basically 
depends on the energy density scale at which the phase transition occurs, what leaves us with just one degree of freedom in order to match this theoretical value with the observational data. This makes of the model a reasonably predictive one. Finally, we have emphasized  that the reheating temperature obtained in our model is the same one gets in inflationary cosmology via gravitational particle production, what clearly shows the dual relation that exists between the matter-ekpyrotic bouncing scenario in Loop Quantum Cosmology and the non-oscillatory inflationary models.

\vspace{1cm}

{\bf Acknowledgements.}
This investigation has been supported in part by MINECO (Spain), projects MTM2011-27739-C04-01, FIS2010-15640 and FIS2013-44881, and by the CPAN Consolider Ingenio Project.

\end{document}